\documentclass{aac}

\usepackage[numbers]{natbib}
\usepackage{rotating}
\usepackage{graphicx}

\usepackage{xcolor}

\begin{document}

\title{Compact Spectral Characterization of 5-500 MeV X-rays from the Texas Petawatt Laser-Driven Plasma Accelerator}

\author[1]{A. Hannasch}
\author[1]{L. Lisi}
\author[1]{J. Brooks}
\author[1]{X. Cheng}
\author[2]{A. Laso Garcia}
\author[1]{M. LaBerge}
\author[1]{I. Pagano}
\author[1]{B. Bowers}
\author[1]{R. Zgadzaj}
\author[1]{H. J. Quevedo}
\author[1]{M. Spinks}
\author[1]{M. E. Donovan}
\author[2]{T. Cowan}
\author[1]{M. C. Downer\corref{cor1}}

\affil[1]{The University of Texas at Austin, Department of Physics, Austin, Texas 78712-1081, USA.}
\affil[2]{The Helmholtz-Zentrum Dresden-Rossendorf, Institute for Radiation Physics, 01328 Dresden, Germany}
\corresp[cor1]{downer@physics.utexas.edu}

\maketitle

\begin{abstract}
We reconstruct spectra of secondary x-rays generated from a 500 MeV - 2 GeV laser plasma electron accelerator. A compact (7.5 $\times$ 7.5 $\times$ 15 cm), modular x-ray calorimeter made of alternating layers of absorbing materials and imaging plates records the single-shot x-ray depth-energy distribution.  X-rays range from few-MeV inverse Compton scattered x-rays to $\sim$100 MeV average bremsstrahlung energies and are characterized individually by the same calorimeter detector. Geant4 simulations of energy deposition from mono-energetic x-rays in the stack generate an energy-vs-depth response matrix for the given stack configuration.  A fast, iterative reconstruction algorithm based on analytic models of inverse Compton scattering and bremsstrahlung photon energy distributions then unfolds x-ray spectra in $\sim10$ seconds.
\end{abstract}

\section{INTRODUCTION}
Accelerator based high energy x-ray sources at photon energies $> 1$ MeV find applications in secondary particle generation \cite{Pomerantz2014UltrashortSource}, material science \cite{Asoka-Kumar1994CharacterizationPositrons,Selim2002DopplerRadiation}, space science \cite{Stassinopoulos1988TheEnvironment}, nuclear resonance fluorescence \cite{Hayward1957PhotonC12,Cohen1959ResonanceB11} and homeland security \cite{Bertozzi2007NuclearContraband}. These x-ray sources require relativistic electron energies, typically from radio-frequency accelerators, to generate bright synchrotron or bremsstrahlung radiation. Radio-frequency accelerator facilities are limited by accelerating gradients of $\sim$100 MeV/m \cite{Allen1989High-GradientKlystron} and require m- to km-scale structures, becoming expensive to build and operate. Laser plasma accelerators (LPAs) offer a tabletop complement to RF accelerators with accelerating gradients $\sim 100$ GeV/m from an intense laser pulse interacting with a plasma \cite{Tajima1979LaserAccelerator,Esarey2009PhysicsAccelerators}.  LPAs have accelerated electrons to energies $E_e$ approaching 10 GeV \cite{Gonsalves2019PetawattWaveguide} with bandwidth $\Delta E_e/E_e\sim 1-15\%$. At lower energies they produce nCs of total charge, enabling them to generate secondary x-rays with peak brilliance comparable to that of synchrotron x-rays.

LPAs can generate three types of secondary x-rays \cite{Corde2013FemtosecondAccelerators}: betatron, inverse Compton scattered (ICS), and bremsstrahlung. Betatron radiation originates from transverse undulations of accelerating electrons in the wake's focusing fields \cite{Esarey2002SynchrotronChannels,Kostyukov2003X-rayChannel} and has a synchrotron-like spectrum with critical energy $\sim$ several 10s of keV from a GeV class LPA \cite{Rousse2004ProductionInteraction}.  ICS radiation results from backscatter of counter-propagating laser photons from accelerated electrons, upshifting the photons to energy $E_x \sim 4 \gamma_e^2 E_L$ \cite{Esarey1993NonlinearPlasmas} where $E_L$ is the laser photon energy and $\gamma_e$ is the Lorentz factor of the electrons. Thus ICS of $E_L = 1.17$ eV photons from electron bunches with peak energy in the range $500 < E_e < 900$ MeV ($980 < \gamma_e < 1760$) generates x-rays with spectral peaks in the range $5 < E_x < 15$ MeV.  Bremsstrahlung x-rays result from collisions of relativistic electrons passing through a converter after the accelerator, producing broadband x-rays with photon energies up to the energy of the electrons \cite{Glinec2005High-resolutionSource}.  Secondary x-ray photons from GeV-class LPAs thus span an energy range from several keV to several hundred MeV, enabling a wide range of applications \cite{Corde2013FemtosecondAccelerators,Albert2014LaserRequirements}.

Here, we spectrally characterize bremsstrahlung and ICS x-rays generated by 500 MeV - 2 GeV LPA electrons in a single shot, using a single, compact, inexpensive calorimeter consisting of a stack of absorbers of varying $Z$ and thickness, interlaced with imaging plates (IPs).  We reconstruct spectra that are bremsstrahlung- or ICS-dominated, and discuss future improvements to enable unfolding of more complex spectra. The calorimeter used here extends earlier work by Hannasch \textit{et al.} (2021) \cite{Hannasch2021CompactAccelerator} to significantly higher x-ray photon energies.

\begin{figure}[ht!]
\centerline{\includegraphics[width=.9\textwidth]{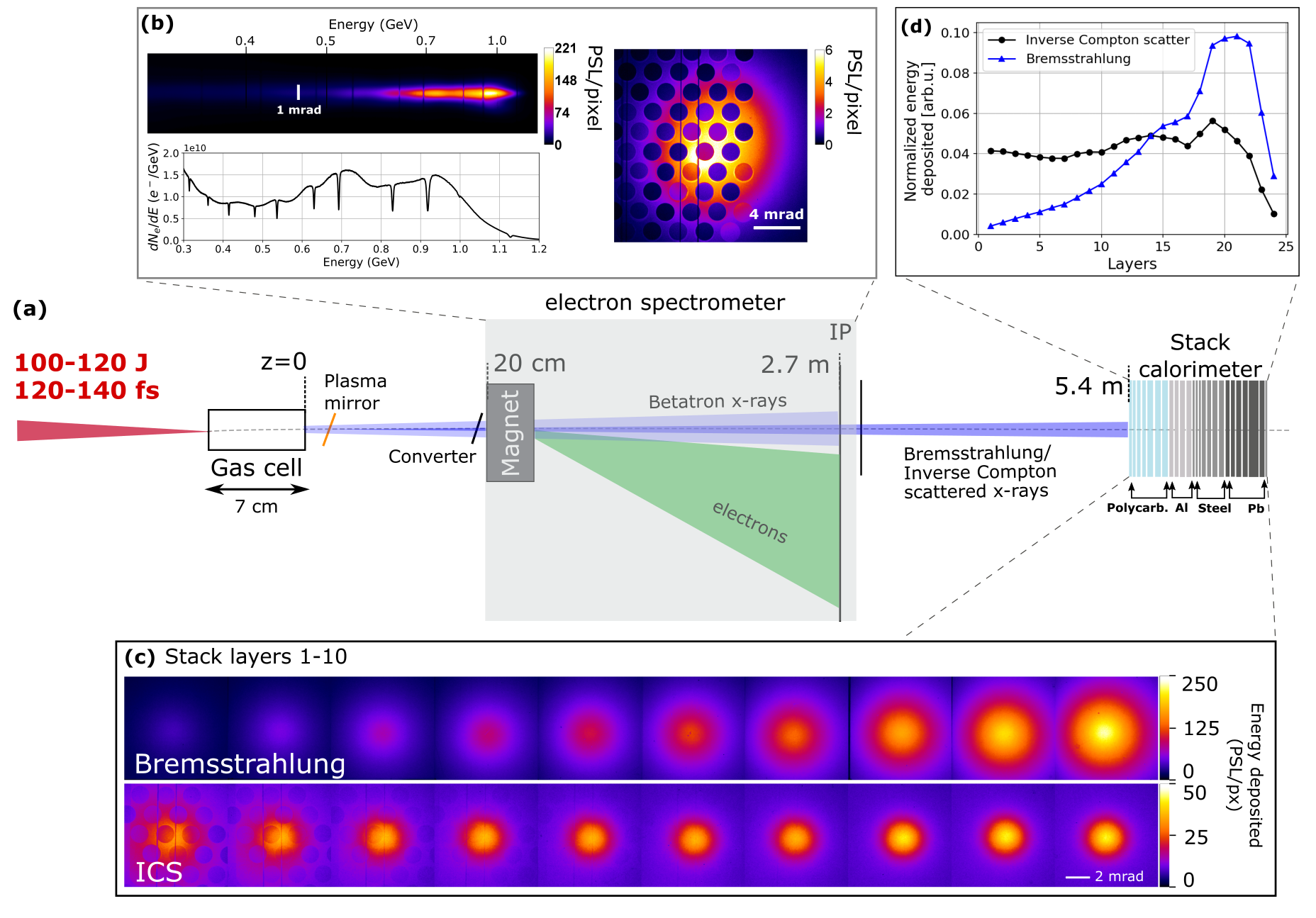}}
\caption{\label{fig:setup} LPA x-ray characterization overview.  (a) Schematic set up showing (left to right) incident laser pulse, gas cell, tilted plasma mirror (PM) positioned at $z=3.3$\,cm from gas cell exit for generating inverse Compton scattered x-rays, converter at $z = 20$\,cm for generating bremsstrahlung, 1 T magnetic electron spectrometer, and stack calorimeter made of alternating absorbers and image plates. (b) Representative single-shot electron spectrum. Top left:  raw dispersed electron data from image plate.  Bottom left:  electron energy distribution averaged over the vertical width of the trace. Right: raw betatron profile after passing through Ross filter pairs. (c) Representative energy-depth profiles from the stack calorimeter image plates for a bremsstrahlung dominated source (top) and an inverse Compton scatter dominated source (bottom). (d) Integrated energy-depth distributions in the stack for the bremsstrahlung (blue triangles) and inverse Compton source (black circles) normalized to the total deposited energy in the stack.}
\end{figure}

\subsection{PROCEDURE}

Experiments were performed at the Texas Petawatt Laser facility at the University of Texas at Austin which delivers 100-120 J pulses, 120-140 fs in duration to a 7 cm long gas cell of pure He at an electron density of $5 \times 10^{17}~\mathrm{cm^{-3}}$ [see Fig.\,\ref{fig:setup}(a)]. A $\sim1$ T magnetic spectrometer placed 20 cm downstream from the gas jet sweeps accelerated electrons off-axis and measures their angle integrated spectrum at an image plate (IP) placed 2.7 m after the exit of the gas cell \cite{Wang2013Quasi-monoenergeticGeV} [see Fig.\,\ref{fig:setup}(b)]. The IP records the on-axis betatron radiation after passing through a Ross-pair filter set for spectral analysis [Fig.\,\ref{fig:setup}(b) right] and wire fiducials placed at two z-positions after the magnet impart shadows on the dispersed electrons which provides energy and pointing calibration \cite{Wang2013Quasi-monoenergeticGeV} [Fig.\,\ref{fig:setup}(b) left, top and bottom]. The bremsstrahlung and ICS x-ray spectra are diagnosed with a compact stack calorimeter made of alternating absorbers of increasing Z and IPs placed 5.4 m after the accelerator and outside of vacuum. Thin target bremsstrahlung x-rays are generated by placing a 25 $\mu$m-thick Kapton film 20 cm after the gas cell and prior to the magnet so that the laser is not intense enough to generate ICS radiation. Thick target bremsstrahlung x-rays are generated by placing a 860 $\mu$m-thick Sn converter prior to the magnet to compare with the 25 $\mu$m-thick Kapton target. ICS dominated radiation is generated by placing the 25 $\mu$m Kapton film 3.3 cm after the exit of the gas cell to act as a plasma mirror and to minimize the bremsstrahlung contribution. We estimate the scattering parameter $a_0<0.5$ when the plasma mirror is placed 3.3 cm after the exit of the accelerator, generating ICS radiation in a linear regime \cite{Corde2013FemtosecondAccelerators}. Figure \ref{fig:setup}(c) shows the first 10 IP exposures for the 860 $\mu$m-thick Sn bremsstrahlung case (top) and the ICS dominated case (bottom). The photostimulated luminescence (PSL) in each IP layer is integrated over the FWHM of the beams and converted to energy using the value $6.95 \pm 1.2 \times 10^{-1}$ PSL/MeV from Bonnet \textit{et al.}, (2013) \cite{Bonnet2013ResponseParticles}. The IP scans were performed 5-20 minutes after the shot and are corrected for fading based on Tanaka \textit{et al.,} (2005) \cite{Tanaka2005CalibrationSpectrometer}. These energy-depth profiles, shown in Fig.\,\ref{fig:setup}(d), are inputs for reconstructing the incident photon spectrum and are normalized to the total deposited energy. The normalized profiles depend only on the relative energy content of the incident photons while the absolute number of photons in the incident x-rays then depends on the unfolded spectra and the energy conversion and fading correction with an estimated combined uncertainty of $\pm 20\%$.

\subsection{X-ray spectral reconstruction}

Reconstruction of the incident x-ray spectrum at the stack follows the procedure and error analysis outlined in Hannasch \textit{et al.} (2021) and is summarized here. The absorbing layers were tailored to maximize resolution for photon energies $> 1$ MeV. The stack design for this experiment consists of polycarbonate, aluminum, steel and lead in increasing thickness. We performed Geant4 simulations of mono-energetic photons interacting with the stack to generate a response matrix $R_{ij}$, which gives the energy deposited in layer \textit{i} by a photon of energy $\hbar \omega_j$. We write the integrated energy deposited in layer $i$ of the calorimeter as a vector with components $D_i ~(i=1,2,...,24)$. $D_i$ can be calculated from the response matrix and a discretized photon spectrum $dN_j/d(\hbar\omega)$ which describes the number of photons of energy $\hbar\omega_j$ in increment $d(\hbar\omega_j)$:

\begin{equation}\label{dose calculation}
    D_i^{(calc)} \approx \sum_{j=1}^{N} \frac{dN_j}{d(\hbar \omega)} R_{ij}  ~d(\hbar \omega_j).
\end{equation}

\noindent
Here, N is the number of energy bins. To unfold the incident photon spectrum, $dN_j/d(\hbar\omega)$ is constrained to take the form of a physics-based analytic function of $\hbar\omega$ that depends on additional parameters of the bremsstrahlung and ICS radiation. We wish to reconstruct the incident photon spectrum $dN_j/d(\hbar\omega)$ by finding these additional parameters that minimize the sum of the squared difference between calculated and measured deposited energy profiles.

\vspace{0.2cm}
\noindent
\textbf{Bremsstrahlung radiation:} The model used for unfolding the bremsstrahlung spectra is based on Kramers' law \cite{Kramers1923OnSpectrum}, where the x-ray spectrum depends on the initial electron energy $E_0$ as 

\begin{equation}\label{Brem model}
    \frac{dN}{d(\hbar \omega)} \propto \frac{1}{E_0}(E_0-\hbar \omega), ~~~~~~~~~ \hbar\omega \leq E_0.
\end{equation}

\noindent
Kramers' law has been used extensively to predict the experimental bremsstrahlung spectra at a given angle \cite{KochH.W.Motz1959BremsstrahlungData} and is a good model for bremsstrahlung from a broadband electron distribution as in the cases presented here. The unfolded parameter, $E_0$, defines the cutoff energy of the x-ray spectrum and is approximately the maximum electron energy for broadband electron spectra. 

\vspace{0.2cm}
\noindent
\textbf{Inverse Compton scattered radiation:} The energy spectrum for \textit{linear} ICS radiation, $dI/d(\hbar\omega)$, can be approximated by a Gaussian function with mean energy $E_x$ and standard deviation of $\sigma_x$ as  

\begin{equation}\label{ICS model}
    \frac{dN}{d(\hbar \omega)} = \frac{1}{\hbar\omega} \frac{dI}{d(\hbar\omega)} \propto \frac{1}{\hbar\omega} \exp\left(-\frac{(\hbar\omega - E_x)^2}{2 \sigma_x^2} \right)
\end{equation}

\noindent
This approximation has proven successful at reconstructing the characteristic features of ICS radiation from peaked electrons with energy spread $>10\%$ and laser strength parameter $a_0 \ll1$, i.e., for linear ICS \cite{Hannasch2021CompactAccelerator}. The unfolded parameters of the energy spectrum are weighted by $(\hbar\omega)^{-1}$ when converting to photon spectrum, thus the peak energy appears to be redshifted from the unfolded $E_{pk}$ while the energy spread appears larger than $\sigma_x$. 


\section{RESULTS}

\subsection{Bremsstrahlung radiation}

\begin{figure}[ht!]
\centerline{\includegraphics[width=.85\textwidth]{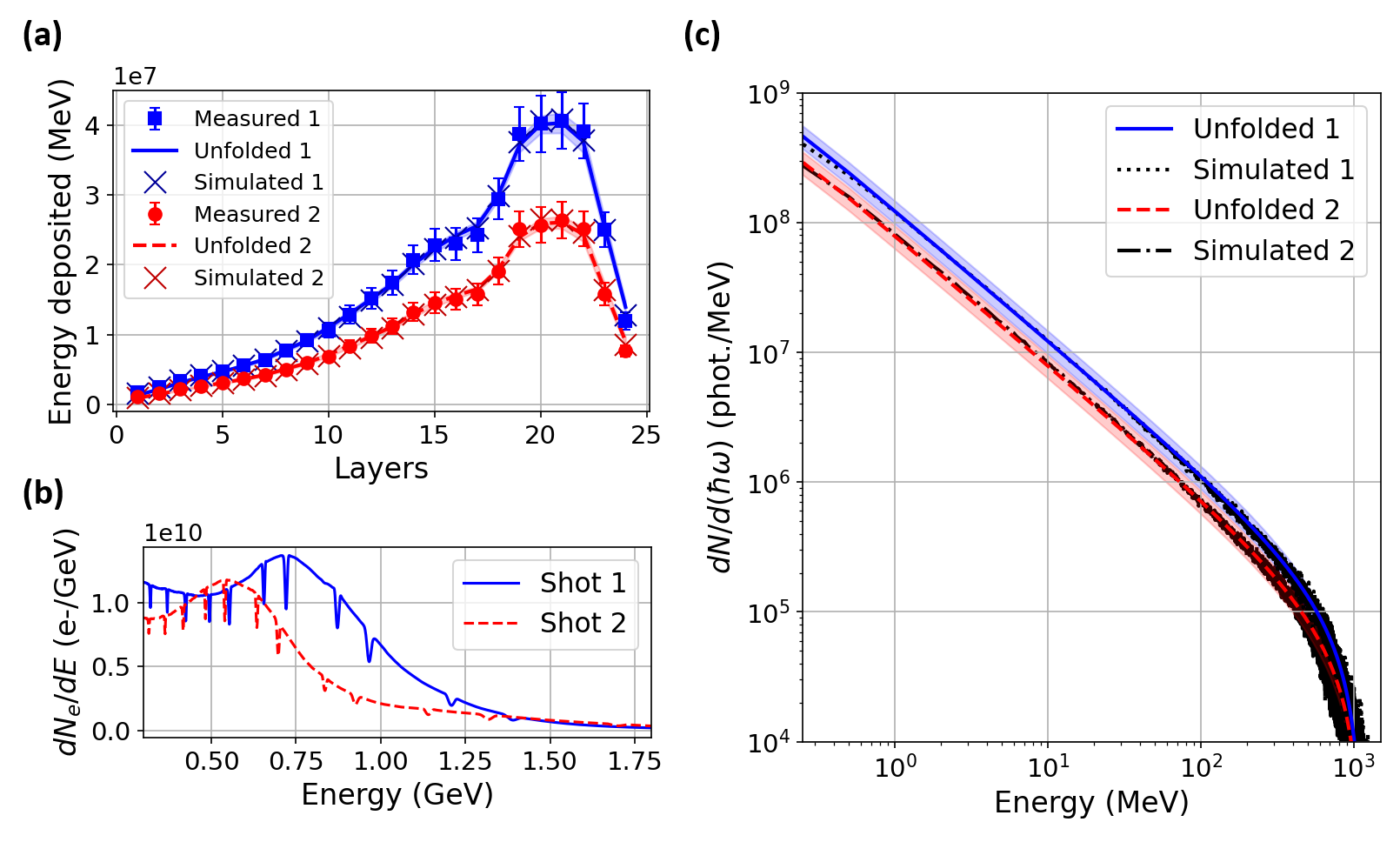}}
\caption{\label{fig:brem plots}860 $\mu$m-thick Sn bremsstrahlung x-ray results. (a) Measured energy deposition profile from shot 1 (blue squares) and shot 2 (red circles), corresponding unfolded profiles (solid blue and dashed red curves) based on bremsstrahlung radiation model (Eq. \ref{Brem model}) and Geant4-simulated profiles ($\times$'s) based on the measured electron spectra in (b); (c) bremsstrahlung spectra extracted from unfolded and simulated profiles in (a) where shaded regions denote unfolding error in (a).}
\end{figure}

The 860 $\mu$m-thick Sn bremsstrahlung integrated energy profiles $D^{(meas)}_i~(i=1,2,... 24)$ for two consecutive shots (1 and 2) are shown in Fig.\,\ref{fig:brem plots}(a) as squares and circles generated from the electron spectra shown in Fig.\,\ref{fig:brem plots}(b). Simulations of the anticipated bremsstrahlung spectra were performed in Geant4 by including the converter material in the Geant4 geometry and initiating electrons with an energy distribution given by the measured electron spectra in Fig.\,\ref{fig:brem plots}(b). Figure \ref{fig:brem plots}(a) compares the measured energy deposition profiles with the unfolded profiles (solid and dashed curves) and simulated profiles ($\times$'s). Figure \ref{fig:brem plots}(c) shows the corresponding unfolded photon spectra (solid and dashed curves) for both shots along with the simulated spectra (dotted and dash-dotted curves) that fall within 10\% of the unfolded spectral shape for each shot. The unfolded and simulated bremsstrahlung parameters are compiled in Table\,\ref{tab:brem parameters}.

\begin{table}[h]
\caption{\label{tab:brem parameters}Compiled 860 $\mu$m-thick Sn bremsstrahlung x-ray results}
\tabcolsep7pt
\begin{tabular}{lccccc}
\hline
  & \tch{1}{c}{b}{Electron $E_{pk}$ \\(MeV)}& \tch{1}{c}{b}{x-ray FWHM \\(mrad) }  & \tch{1}{c}{b}{Unfolded $E_{avg}$ \\(MeV)}  & \tch{1}{c}{b}{Simulated $E_{avg}$ \\(MeV)}  & \textbf{Unfolded $N_{phot}$}    \\
\hline
Shot 1  & $680 \pm 280$ & $4.0 \pm 0.1$ & $84 \pm 6$ &  86 & $7.9 \pm 1.6\times 10^8$ \\
\hline
Shot 2  & $530 \pm 220$ & $4.0 \pm 0.1$ & $85 \pm 7$ & 83 & $5.1 \pm 1.0 \times 10^8$ \\
\hline
\end{tabular}
\end{table}

\begin{figure}[ht!]
\centerline{\includegraphics[width=.85\textwidth]{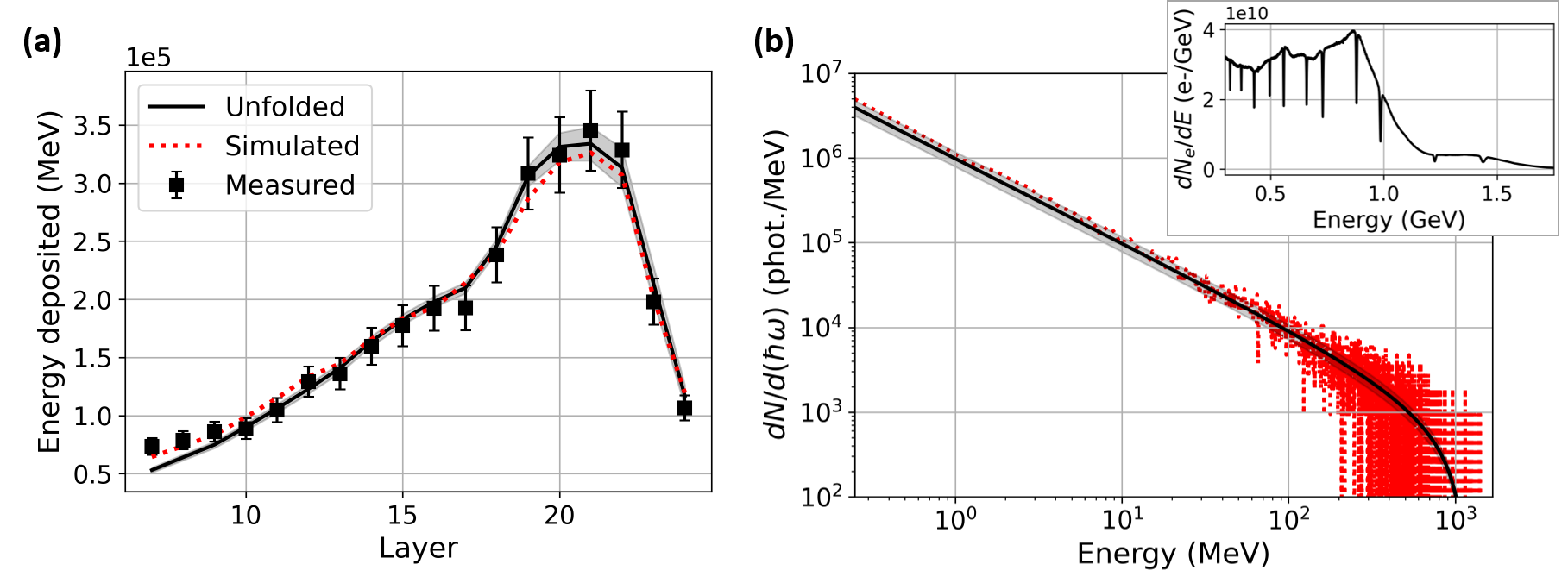}}
\caption{\label{fig:kapton brem}25 $\mu$m-thick Kapton bremsstrahlung x-ray results. (a) Measured energy deposition profile (black squares), unfolded profile (black solid curve) based on the bremsstrahlung radiation model (Eq. \ref{Brem model}) and Geant4-simulated profile (red dotted curve) based on the measured electron spectra in inset of (b); (b) bremsstrahlung spectra extracted from unfolded and simulated profiles in (a) where shaded regions denote unfolding error in (a).}
\end{figure}

The 25 $\mu$m-thick Kapton bremsstrahlung integrated energy profile $D^{(meas)}_i~(i=7,8,... 24)$ is shown in Fig.\,\ref{fig:kapton brem}(a) as data points. Betatron radiation that is not blocked by the Kapton deposits energy in layers 1-6 as evident from the 2D pattern of the Ross pair filters used to analyze their spectrum observed in those layers [see Fig.\,\ref{fig:setup}(c) bottom panel, left-most layer images]. These layers are removed for unfolding the bremsstrahlung only signal from 25 $\mu$m-thick Kapton. The dotted red and solid black curves in Fig.\,\ref{fig:kapton brem}(a) show the simulated and unfolded deposited energy profiles and the corresponding simulated and unfolded spectra are shown in Fig.\,\ref{fig:kapton brem}(b). The simulated average photon energy is 76 MeV and falls within the uncertainty of the unfolded average energy of $81 \pm 8$ MeV. The bremsstrahlung from the 25 $\mu$m-thick Kapton generates $3.5 \pm 0.7 \times 10^6$ photons within the FWHM, $\sim 200 \times$ fewer photons than is generated by the 860 $\mu$m-thick Sn converter. 

\subsection{Inverse Compton scattered radiation}

\begin{figure}[ht!]
\centerline{\includegraphics[width=.85\textwidth]{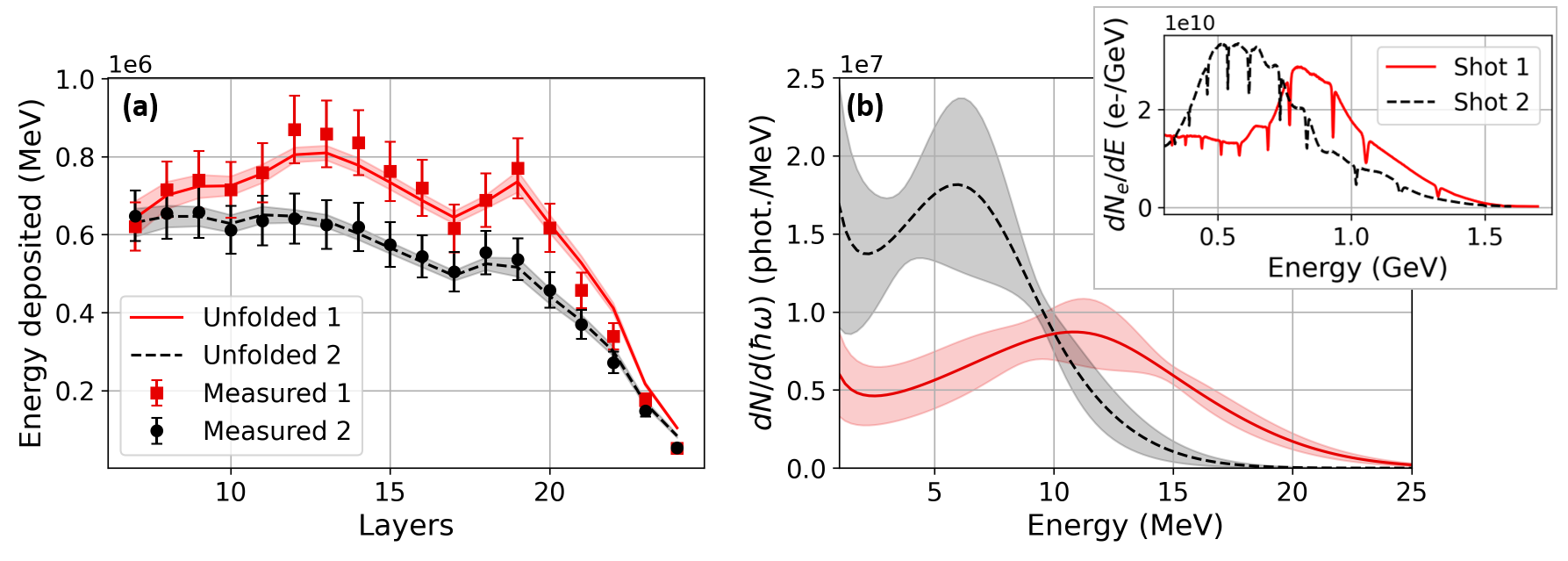}}
\caption{\label{fig:ICS plots}ICS x-ray results. (a) Measured energy deposition profile from shot 1 (red squares) and shot 2 (black circles) generated by the electrons spectra in the inset of (b) and unfolded profiles (solid and dashed curves) based on the ICS radiation model (Eq. \ref{ICS model}); (b) ICS spectra extracted from unfolded profiles in (a) where shaded regions denote unfolding error in (a).}
\end{figure}

The ICS integrated energy profiles $D^{(meas)}_i~(i=7,8,... 24)$ for two consecutive shots (1 and 2) are shown in Fig.\,\ref{fig:ICS plots}(a) as squares and circles generated from the electron spectra shown in the inset plot of Fig.\,\ref{fig:ICS plots}(b). The first 6 layers are discarded to remove the betatron contribution and Ross pair filter shadows which are evident in Fig.\,\ref{fig:setup}(c) in the bottom panel. The bremsstrahlung contribution from the 25 $\mu$m-thick Kapton is scaled to the total energy of the electron beam $N_e\left< E_e\right>$ for each shot and subtracted from $D_i^{(meas)}$ to unfold the ICS \textit{only} contribution. The unfolded deposited energy profiles are overlayed in Fig.\,\ref{fig:ICS plots}(a) as solid and dashed curves and the corresponding unfolded photon spectra are shown in Fig.\,\ref{fig:ICS plots}(b). The electron peak energy and unfolded x-ray parameters are compiled in Table \,\ref{tab:ICS parameters}. The expected $4\gamma_e \hbar\omega_L$ peak x-ray energy is $12.5 \pm 3$ MeV and $6.3 \pm 2.4$ MeV and falls within the uncertainty of the unfolded $E_{pk}$ for shot 1 and 2, respectively. The ICS shots generate $\sim10\times$ fewer photons than the 860 $\mu$m-thick Sn bremsstrahlung and $\sim20\times$ more than the 25 $\mu$m-thick Kapton bremsstrahlung.

\begin{table}[h!]
\caption{\label{tab:ICS parameters}Compiled ICS x-ray results}
\tabcolsep7pt
\begin{tabular}{lccccc}
\hline
  & \tch{1}{c}{b}{Electron $E_{pk}$ \\(MeV) }& \tch{1}{c}{b}{x-ray FWHM \\(mrad) }  & \tch{1}{c}{b}{Unfolded $E_{pk}$ \\(MeV)}  & \tch{1}{c}{b}{Unfolded $\sigma_{x}$ \\(MeV)}  & \textbf{Unfolded $N_{phot}$}    \\
\hline
Shot 1  & $840 \pm 230$ & $2.35 \pm 0.05$ & $12.8 \pm 0.8$ &  $4.9 \pm 0.4$ & $6.4 \pm 1.2 \times 10^7$ \\
\hline
Shot 2  & $590 \pm 230$ & $2.55 \pm 0.05$ & $7.9 \pm 0.5$ & $3.6 \pm 0.7$ & $9.1 \pm 1.8 \times 10^7$ \\
\hline
\end{tabular}
\end{table}

\section{DISCUSSION}

The stack calorimeter can unfold the average or peak photon energy and integrated photon number for both the broadband bremsstrahlung and linear inverse Compton scattered radiation sources based on a parameterized model of each source. More complex sources such as linear ICS radiation from electron bunches with multiple energy peaks or nonlinear ICS radiation with multiple harmonics may require a more elaborate unfolding procedure. 

One way to increase the accuracy of the unfolding is to take advantage of material transitions in the stack design. The dips and subsequent increases in the integrated energy observed in the ICS profiles in Fig.\,\ref{fig:ICS plots}(a) are correlated with transitions from low to high Z materials. For example, IP layer 11 is placed between the last Al layer and the first steel layer and absorbs more energy due to back scattering from the higher Z layer. This occurs again as the absorbers transition from steel to lead at IP layer 18. These wiggles in the energy deposition profile for energies between 5 and 10 MeV improve the accuracy of the unfolded spectrum, but are washed out by energetically broader bremsstrahlung x-rays [see Fig.\,\ref{fig:setup}(d), Fig.\,\ref{fig:brem plots}(a) and Fig.\,\ref{fig:kapton brem}(a)]. 

\begin{figure}[ht!]
\centerline{\includegraphics[width=.9\textwidth]{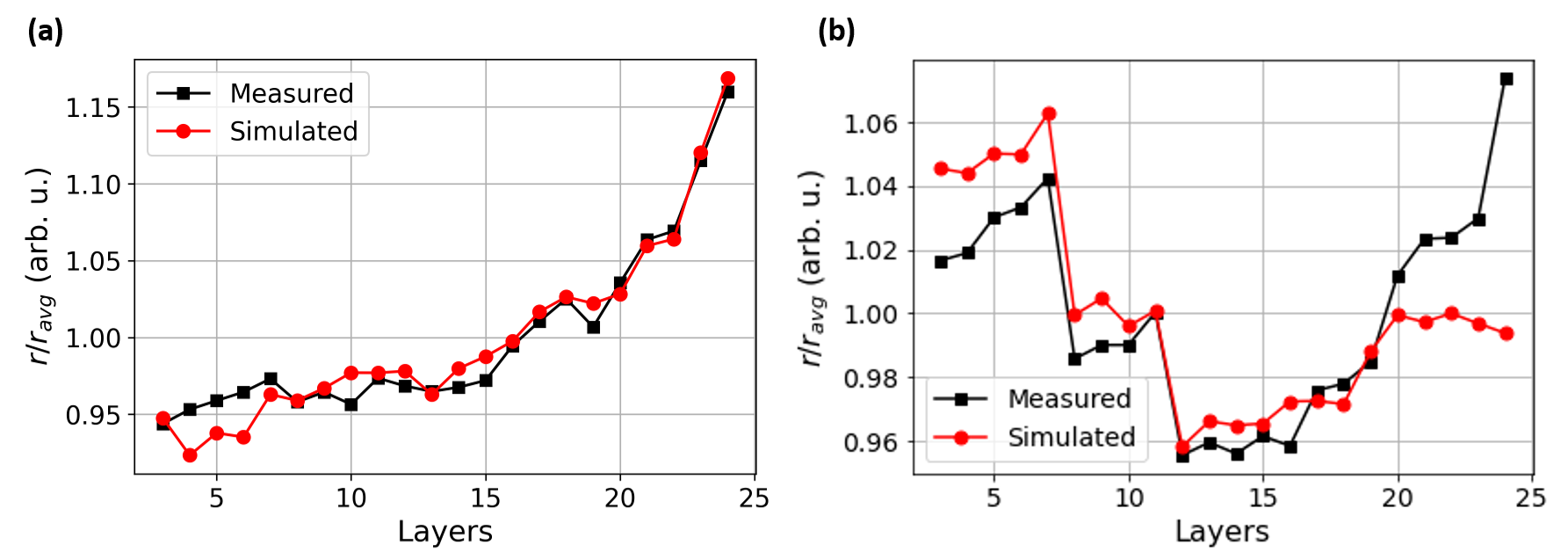}}
\caption{\label{fig:radial plots} (a) The normalized beam radius $r/r_{avg}$ from shot 1 of the 860 $\mu$m-thick Sn bremsstrahlung comparing measured (black squares) and simulated (red circles) data and (b) the normalized beam radius from shot 2 of the ICS (black squares) and 5 MeV simulated mono-energetic data (red circles).}
\end{figure}

Another feature of the deposited energy profile that correlates with transitions from low to high Z materials is the transverse profile size. Similar to the energy deposition for bremsstrahlung, the beam size of a broadband source of x-rays $E_{avg} \gg 1$ MeV will experience a relatively smooth and constant increase through the stack. In contrast, a peaked source $\sim 5$ MeV such as ICS experiences sharp changes in the deposited energy profile radius at the material transitions. Figure \,\ref{fig:radial plots}(a) shows the measured and simulated normalized radius $r/r_{avg}$ as the 860 $\mu$m-thick Sn bremsstrahlung radiation passes through the stack. The same normalized radius is shown in Fig.\,\ref{fig:radial plots}(b) for shot 2 of the unfolded ICS source ($E_{pk} = 7.9 \pm 0.5$) and a simulated mono-energetic beam at 5 MeV. These sharp features in the radiation profile radius as a function of layer in the stack can be a powerful tool to aid unfolding, reducing the number of possible solutions that may match the energy deposition profile.

\section{CONCLUSION}

We have presented a set of unfolded secondary x-ray spectra from LPA electrons with energies between 500 MeV and 900 MeV. Bremsstrahlung from an 860 $\mu$m Sn target is unfolded from two electron spectra with average energies of $84\pm 6$ MeV and $85 \pm 7$ MeV and with $7.9\pm1.6\times10^8$ and $5.1\pm1.0\times10^8$ photons within the FWHM. These bremsstrahlung cases are compared with Geant4 simulations of the anticipated spectrum and agree well within the uncertainty.  Thin-target bremsstrahlung from $25~\mu$m-thick Kapton is unfolded with an average energy is $81\pm 8$ MeV and $3.6 \pm 0.7 \times 10^ 6$ photons in the FWHM. ICS dominated radiation from electron bunches with different peak energies was unfolded to observe a shift in peak ICS energy from $7.9\pm0.5$ MeV to $12.8\pm 0.8$ MeV and a total of $9.1\pm1.8\times10^{7}$ and $6.4\pm1.2\times10^{7}$ photons in the FWHM, respectively. The versatility of the stack calorimeter makes it a valuable diagnostic for LPA based x-ray sources over a large range of photon energies.

\section{ACKNOWLEDGEMENTS}
U. Texas authors acknowledge support from the U.S. Department of Energy grant DE-SC0011617 and the DOE Office of Science, Fusion Energy Sciences under Contract No. DE-SC0019167 through the LaserNetUS initiative at the University of Texas at Austin. A. H. acknowledges support from the National Science Foundation Graduate Research Fellowship under Grant No. DGE-1610403.

\section*{AUTHOR CONTRIBUTIONS STATEMENT}
A.H., L.L., J.B., X.C., A.L.G., M.L., I.P., and B.B. conducted the experiments. A.H. analyzed the results. A.H., A.L.G. and L.L. performed the simulations. H.J.Q., M.S.,and M.E.D. manage and operate the Texas Petawatt laser facility. R.Z., T.C., and M.C.D. provided overall supervision of the project. All authors reviewed the manuscript. 

\bibliographystyle{aac}%
\bibliography{references_static}%

\end{document}